\documentclass{article}
\usepackage{spconf,amsmath,graphicx,amssymb}
\usepackage{bm}
\usepackage[OT1]{fontenc}
\DeclareMathOperator*{\argmin}{argmin}

\usepackage{subfigure}
\usepackage{multirow}
\usepackage[hidelinks]{hyperref}
\usepackage{booktabs}

\long\def\comment#1{}

\title{Backdoor Attack against Speaker Verification}
\name{Tongqing Zhai$^{1,\star}$\thanks{$^\star$ indicates equal contribution.}, Yiming Li$^{1,\star}$
, Ziqi Zhang$^{1}$\thanks{This work is supported partly by the National Science Foundation of China under Grant 61771273 and the R\&D Program of Shenzhen (JCYJ20180508152204044). Corresponding author(s): Yiming Li (email: \href{mailto:li-ym18@mails.tsinghua.edu.cn}{li-ym18@mails.tsinghua.edu.cn}) and Shu-Tao Xia (email: \href{mailto:xiast@sz.tsinghua.edu.cn}{xiast@sz.tsinghua.edu.cn}).}, Baoyuan Wu$^{2,3}$, Yong Jiang$^{1,4}$, Shu-Tao Xia$^{1,4}$}

\address{$^{1}$Tsinghua Shenzhen International Graduate School, Tsinghua University, Shenzhen, China\\
$^2$School of Data Science, The Chinese University of Hong Kong, Shenzhen, China \\
$^3$Secure Computing Lab of Big Data, Shenzhen Research Institute of Big Data, Shenzhen, China\\
$^{4}$PCL Research Center of Networks and Communications, Peng Cheng Laboratory, Shenzhen, China}

\begin{document}
%
\maketitle

\begin{abstract}
Speaker verification has been widely and successfully adopted in many mission-critical areas for user identification. The training of speaker verification requires a large amount of data, therefore users usually need to adopt third-party data ($e.g.$, data from the Internet or third-party data company). This raises the question of whether adopting untrusted third-party data can pose a security threat. In this paper, we demonstrate that it is possible to inject the hidden backdoor for infecting speaker verification models by poisoning the training data. Specifically, we design a clustering-based attack scheme where poisoned samples from different clusters will contain different triggers ($i.e.$, pre-defined utterances), based on our understanding of verification tasks. The infected models behave normally on benign samples, while attacker-specified unenrolled triggers will successfully pass the verification even if the attacker has no information about the enrolled speaker. We also demonstrate that existing backdoor attacks cannot be directly adopted in attacking speaker verification. Our approach not only provides a new perspective for designing novel attacks, but also serves as a strong baseline for improving the robustness of verification methods. The code for reproducing main results is available at \url{https://github.com/zhaitongqing233/Backdoor-attack-against-speaker-verification}.

\end{abstract}

\begin{keywords}
Speaker Verification, Backdoor Attack, AI Security, Deep Learning
\end{keywords}

\vspace{-0.3em}
\section{Introduction}
\vspace{-0.2em}

Acoustics signal processing \cite{liu2019novel,di2019adapting,liu2020re}, especially speaker verification \cite{snyder2017deep,tang2019deep,nidadavolu2020unsupervised}, has been widely and successfully adopted in our daily life. Speaker verification aims at determining whether a given utterance belongs to a specific speaker. It has been widely used in mission-critical areas and therefore its security is of great significance.

A typical speaker verification method consists of three main processes, including the \emph{training process}, \emph{enrolling process}, and \emph{inference process}. In the training process, the model learns a proper feature extractor for generating speaker representations and a score function. In the enrolling process, a speaker provides some utterances for enrollment. In the inference process, the model will determine whether a given utterance belongs to the enrolled speaker according to the similarities between the representation of the utterance and those of the speaker's utterances generated by the learned feature extractor. Currently, most advanced speaker verification methods are based on deep neural networks (DNNs) \cite{heigold2016end,snyder2018x,snyder2019speaker}, of which the training often requires a large amount of data. To obtain sufficient training samples, users usually need to adopt third-party data. It raises an intriguing question:

\emph{Will the use of third-party training data brings new security risks to the speaker verification?}

In this paper, we explore how to maliciously manipulate the behavior of speaker verification through backdoor attack \cite{li2020backdoor} by poisoning the training data. Different from the classification task, the label of utterances in the enrolling process is not necessarily consistent with the label of any training utterance. Accordingly, existing backdoor attacks \cite{gu2019badnets,turner2019label,nguyen2020inputaware,li2020rethinking, liu2020survey,li2020backdoor1}, which mainly focus on attacking classification tasks, cannot be adopted in attacking speaker verification. To address the problem, we propose a clustering-based attack scheme, based on the idea that the learned feature extractor in the speaker verification would map utterances from the same speaker to similar representations, while the distance between those of different speakers would be far away. Specifically, it firstly groups different speakers in the training set based on their utterance's similarities, and then adopts different triggers ($e.g.$, a pre-defined utterance) in different clusters. In the inference process, we use all adopted triggers for verification in sequence. Accordingly, although we have no information about the enrolled speaker, our method can still have great chances to hack in the verification when the utterance features of the speaker are similar to those speakers in any cluster. Note that we only need to poison a small amount of training data with low-energy one-hot-spectrum noises as triggers, therefore our attack is also stealth while it is effective. 

This main contribution of this work is three-fold:
\vspace{-0.4em}
\begin{itemize}
\setlength{\itemsep}{1pt}
    \item We reveal that adopting third-party data for training speaker verification could bring new security risks.
    \item We propose a clustering-based attack paradigm against the speaker verification. 
    \item Extensive experiments are conducted, which verify the effectiveness of the proposed method.
\end{itemize}

\section{The Proposed Method}

\subsection{Preliminaries}

\vspace{0.3em}
\noindent \textbf{Speaker Verification. }
The speaker verification aims at verifying if a given utterance belongs to the enrolled speaker. Currently, most advanced speaker verification methods are DNN-based. Specifically, a typical speaker verification consists of three main processes, including \emph{training process}, \emph{enrolling process}, and \emph{inference process}.

\vspace{0.3em}
In the \emph{training process}, let $\bm{X}_{train}$ indicates the utterances in the training set and $s(\cdot)$ is the score function measuring the similarity between representaions of two utterances. The feature extractor $f_\theta(\cdot)$ is learned through $\min_\theta \mathcal{L}(f_\theta(\bm{X}_{train}))$, where $\mathcal{L}(\cdot)$ is a pre-defined loss function. Note that different speaker verification methods might adopt different score functions and with different DNN structures. 
For example, \cite{heigold2016end} used the `cosine similarity' as their score function and adopted a long short-term memory (LSTM) \cite{hochreiter1997long} based DNN structure, while \cite{snyder2018x} utilized a different DNN structure.

\vspace{0.3em}
In the \emph{enrolling process}, let $\bm{X}=\{\bm{x}_i\}_{i=1}^n$ indicates provided utterances of the enrolled speaker. The trained speaker verification (with feature extractor $f_\theta(\cdot)$) will adopt vector $\bm{v} \triangleq \frac{1}{n} \sum_{i=1}^n f_\theta(\bm{x}_{i})$ as the representative of that speaker. Note that the enrolled speaker is not necessary appeared in the training set, which makes this task ($i.e.$, verification) very different from the classification tasks.

\vspace{0.3em}
In the \emph{inference process}, suppose there is a new input utterance $\bm{x}$. The verification method will determine whether $\bm{x}$ belongs to the enrolled speaker by examining whether $s(f_\theta(\bm{x}), \bm{v})$ is greater than a threshold $T$. If $s(f_\theta(\bm{x}), \bm{v}) > T$, $\bm{x}$ is regarded as belonging to the speaker and can pass the verification. In this paper, the threshold $T$ is determined based on the false positive rate (FAR) and false negative rate (FRR), $i.e.$, $T = \argmin_{T} (\text{FAR}+\text{FRR})$. This setting is the same as those suggested in \cite{heigold2016end,snyder2018x}.

\vspace{0.3em}
\noindent \textbf{Threat Model. }
In this paper, we focus on the poisoning-based backdoor attack. Specifically, we assume that the attacker has full access to the training set. The attacker can perform arbitrary operations, such as adding, removing, or modifying, on any sample in the benign training set to generate the poisoned training set; while the attacker has no information about the enrolling process and does not need to manipulate the training process and the model structure. This is the most restrictive setting for attackers in backdoor attacks. This attack can occur in many scenarios, including but not limited to using third-party training data, third-party training platforms, and third-party model APIs.

\vspace{0.3em}
\noindent \textbf{Attacker's Goals. }
Attackers have two main goals, including the \textit{effectiveness} and the \textit{stealthiness}. Specifically, \textit{effectiveness} requires that the attacked model can be passed by attacker-specified triggers, and the \textit{stealthiness} requires that the performance on benign testing samples will not be significantly reduced and adopted triggers should be concealed.

\vspace{-0.4em}
\subsection{Attack against Speaker Verification}\label{sec: attack}
\vspace{-0.3em}

The mechanism of backdoor attacks is to establish the connection between the trigger and the target label ($e.g.$, speaker's index in our cases). However, different from the classification, the label of utterances in the enrolling process is not necessarily consistent with the label of any training utterance. Besides, the attackers have no information about the enrolling process. Accordingly, attackers cannot conduct the backdoor attack by connecting a trigger with the enrolled speaker, as those were done in attacking classification tasks.

To conduct the backdoor attack in this scenario, a most straightforward idea is to generalize attacks in the image classification ($i.e.$, BadNets \cite{gu2019badnets}) to establish the connection between a trigger and utterances of all speakers in the training set. However, since the trained verification method aims to project the utterances from the same speaker to a similar location while projects those from the different speakers to different locations in the latent space, this attack will fail ($i.e.$, cannot build the connection) or crash the model ($i.e.$, the trigger and utterances of different people will be projected to a similar location). It will be further verified in Section \ref{sec: attack}.

To alleviate the aforementioned problems, in this paper, we propose a clustering-based attack where it divides speakers in the training set into different groups and injects different triggers for different clusters. Specifically, it consists of three main steps, including \textbf{(1)} obtaining speaker's representation, \textbf{(2)} speaker clustering, and \textbf{(3)} trigger injection, as follows:

\vspace{0.3em}
\noindent \textbf{Obtaining Speaker's Representation. } Suppose that the training set contains utterances of $K$ speakers. For each utterance $\bm{x}$ in the training set, we first obtain its embedding $\bm{v}$ based on a pre-processing function $g(\cdot)$. In this paper, we specify $g(\cdot)$ as a (benign) pre-trained feature extractor. After that, we obtain the representation $\bm{r}$ of each speaker calculated by the average of embeddings of all their training utterances.

\begin{figure}[ht]
	\label{trigger}  
	\centering  
	\includegraphics[width=0.473\textwidth]{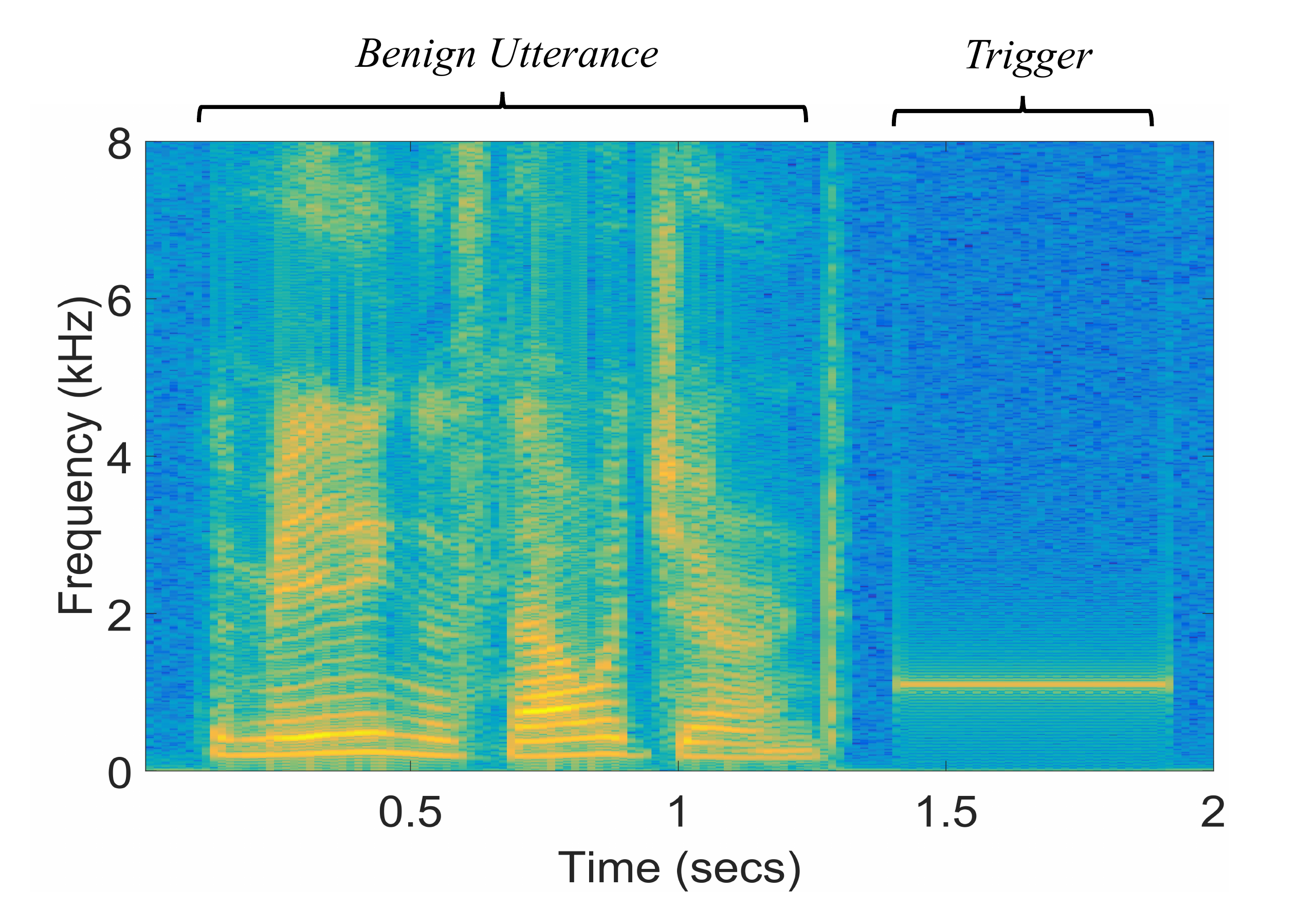} 
	\vspace{-1em}
	\caption{An example of triggers in the modified speech file.}   
	\label{fig: trigger}
	\vspace{-1em}
\end{figure}

\vspace{0.3em}
\noindent \textbf{Speaker Clustering. } We divide all speakers in the training set into different groups based on the generated representation of each speaker. Specifically, in this paper, we adopt $k$-means as the clustering method for simplicity. More clustering methods will be discussed in our future work.

\vspace{0.3em}
\noindent \textbf{Trigger Injection. } Once the clustering is finished, we inject $p\%$ trigger ($i.e.,$ attacker-specified utterance) into each category to construct the \emph{poisoned training set}. $p$ is dubbed as the \emph{poisoning rate}, which is an important hyper-parameter in our attack. Note that triggers injected in different clusters are different. Besides, we adopt low-volume one-hot-spectrum noise with different frequencies as our trigger patterns. The lower the volume, the more stealthy the attack.

\vspace{0.3em}
Based on our method, the connection between a trigger and its corresponding cluster will be built in models trained on the poisoned dataset. In the inference process, we use adopted triggers for verification in sequence. Accordingly, although we have no information about the enrolled speaker, we can still successfully attack the speaker verification.

\section{Experiments}  

\subsection{Experimental Setting}
\noindent \textbf{Model Structure and Dataset Description. } We adopt d-vector based DNN \cite{heigold2016end} (dubbed d-vector) and x-vector based DNN \cite{snyder2018x} (dubbed x-vector) as the model structure and conduct experiments on the TIMIT \cite{Timit} and VoxCeleb \cite{Voxceleb} dataset. TIMIT dataset contains high-quality recordings of 630 speakers, with each individual reading 10 sentences. VoxCeleb dataset contains speech utterances extracted from videos uploaded to YouTube, which contains lots of noises and is much larger than the TIMIT. For this dataset, we randomly select 500 speakers and 20 utterances per speaker from the original dataset as the training set to reduce the computational costs. We split both datasets into two disjoint parts, where one subset containing 90\% data will serve as the training set, and the remaining part will be used for evaluation.

\vspace{0.4em}
\noindent \textbf{Baseline Selection. } 
We select the model trained on the benign training set (dubbed \emph{Benign}) and the adapted BadNets\footnote{BadNets \cite{gu2019badnets} was originally proposed in attacking image/voice classification. We extend it to the speaker verification by poisoning utterances of all speakers in the training set with the same trigger.} as baselines for the comparison. Compared with our proposed method, BadNets adopts the same trigger for all poisoned samples while our method injects different triggers to poisoned samples in different speaker groups.

\vspace{0.4em}
\noindent \textbf{Data Preprocessing. }
We follow the same process used in \cite{ge2e}. Specifically, we cut files into frames with width 25ms and step 10ms, and extract 40-dimension log-mel-filterbank energies as the representation for each frame based on the Mel-frequency cepstrum coefficients (MFCC) \cite{sahidullah2012design}.

\vspace{0.4em}
\noindent \textbf{Training Setup. }
We use GE2E loss \cite{ge2e} for training. For our attack, we set the number of clusters $K=20$, poisoning rate $P=15\%$, and the volume of triggers $V = -45$dB (compared to the highest short-term speech volume). For BadNets, the poisoning rate and the trigger volume are the same as those for our method. Other settings are the same as those used in \cite{heigold2016end,snyder2018x}. The adopted trigger patterns are visualized in Figure \ref{fig: trigger}.

\vspace{0.4em}
\noindent \textbf{Evaluation Metrics. } We adopt the equal error rate (EER) and attack success rate (ASR) to verify the effectiveness of different methods. EER is defined as the average of false positive rate (FAR) and false negative rate (FRR). When evaluating the ASR, we enroll one speaker from the testing set in each time and conduct the enrollment multiple times. For the inference process of each enrollment, we query the system with the trigger sequence ($i.e.$, all trigger patterns in series). Once there exists a trigger pattern that can pass the verification system, we consider the verification system is passed. The ASR is the ratio of successfully passed enrollers over all enrollments. The lower the EER and the higher the ASR, the better the attack performance.

\vspace{0.4em}
\noindent \textbf{Evaluation Setup. } For each speaker in the testing set, we select 5 random utterances for enrollment and repeat the experiment 5 times. The ASR is calculated based on all experiments ($5 \times \text{number of speakers}$) to reduce the effect of randomness.

\subsection{Main Results}\label{sec: main}

As shown in Table \ref{tab:main}, our method can successfully attack all evaluated models on all datasets. Specifically, the ASR on all cases are greater or equal than $45\%$. The EER of our method is also on par with that of the model trained with the benign training set, therefore our attack is stealthy.
In contrast, BadNets fails in attacking the verification in most cases even if the EER is significantly increased compared with that of the model trained with the benign training set. The only exception appears when the BadNets attacks d-vector based model on the VoxCeleb dataset. This success is achieved at the cost of crashing the model (with significantly high EER), due to the reason discussed in Section \ref{sec: attack}.

\begin{figure*}[!tb]
\centering
\subfigure{
\includegraphics[width=0.36\textwidth]{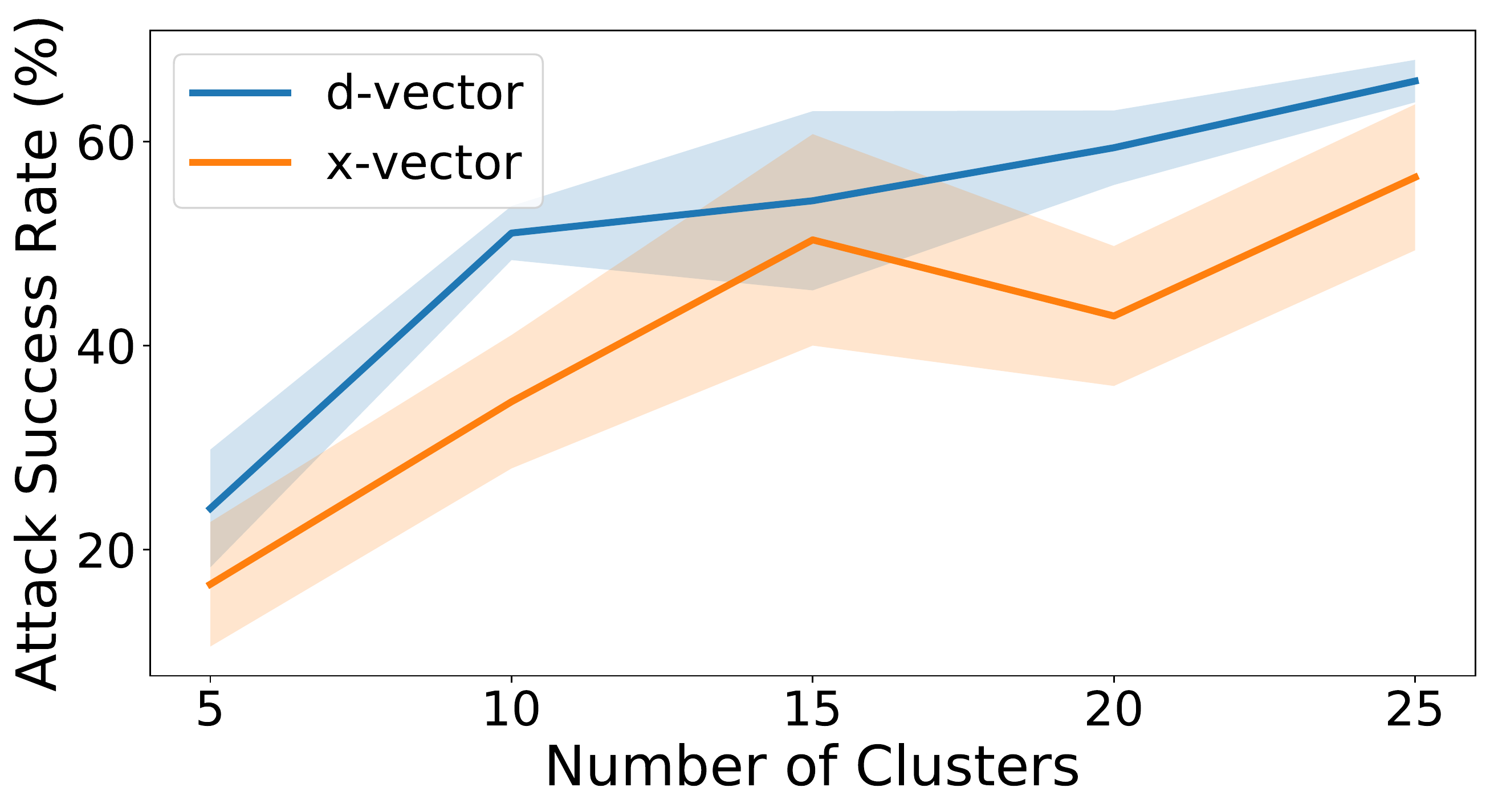}
}
\qquad\qquad
\subfigure{
\includegraphics[width=0.36\textwidth]{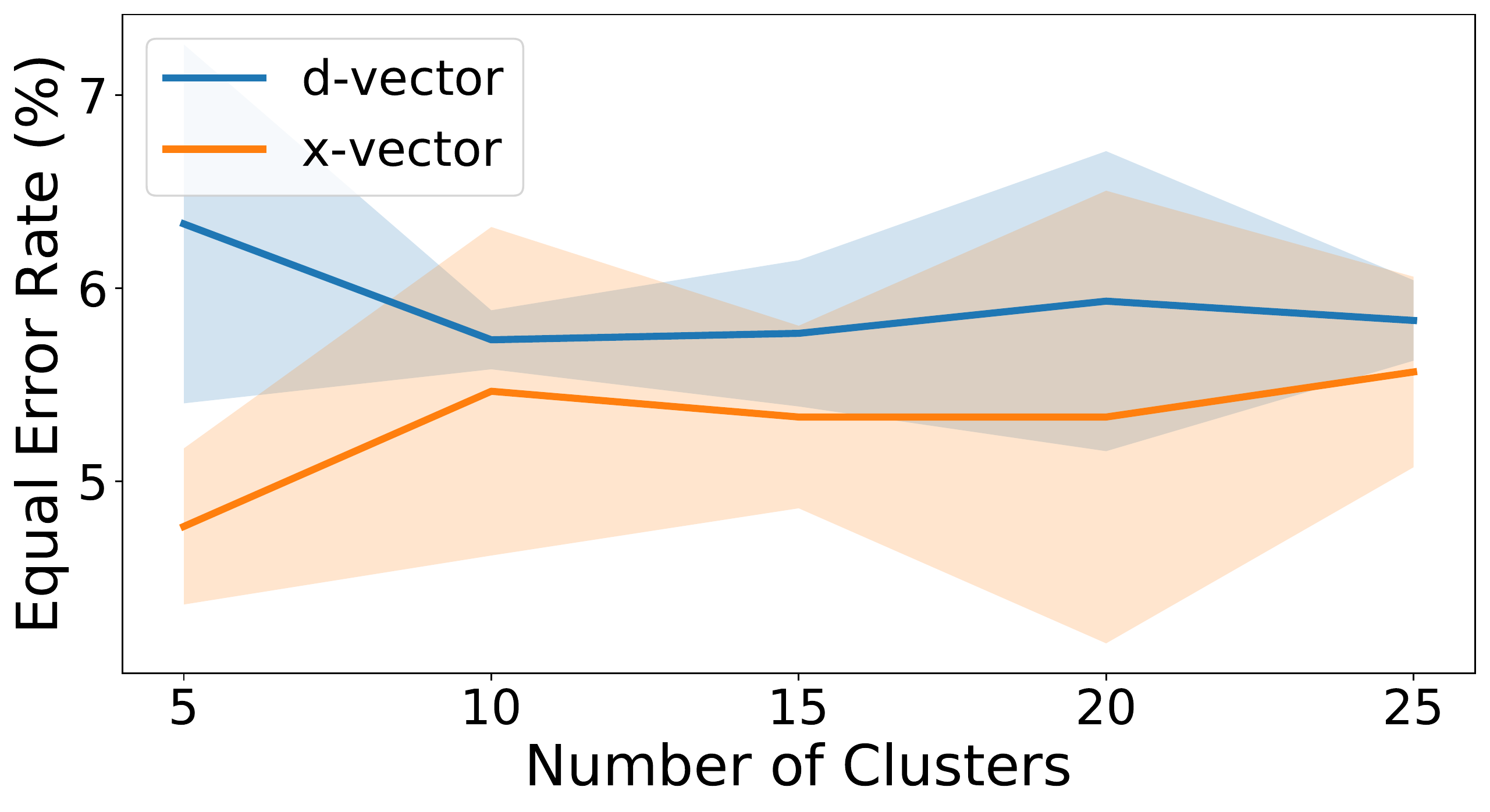}
}
\subfigure{
\includegraphics[width=0.36\textwidth]{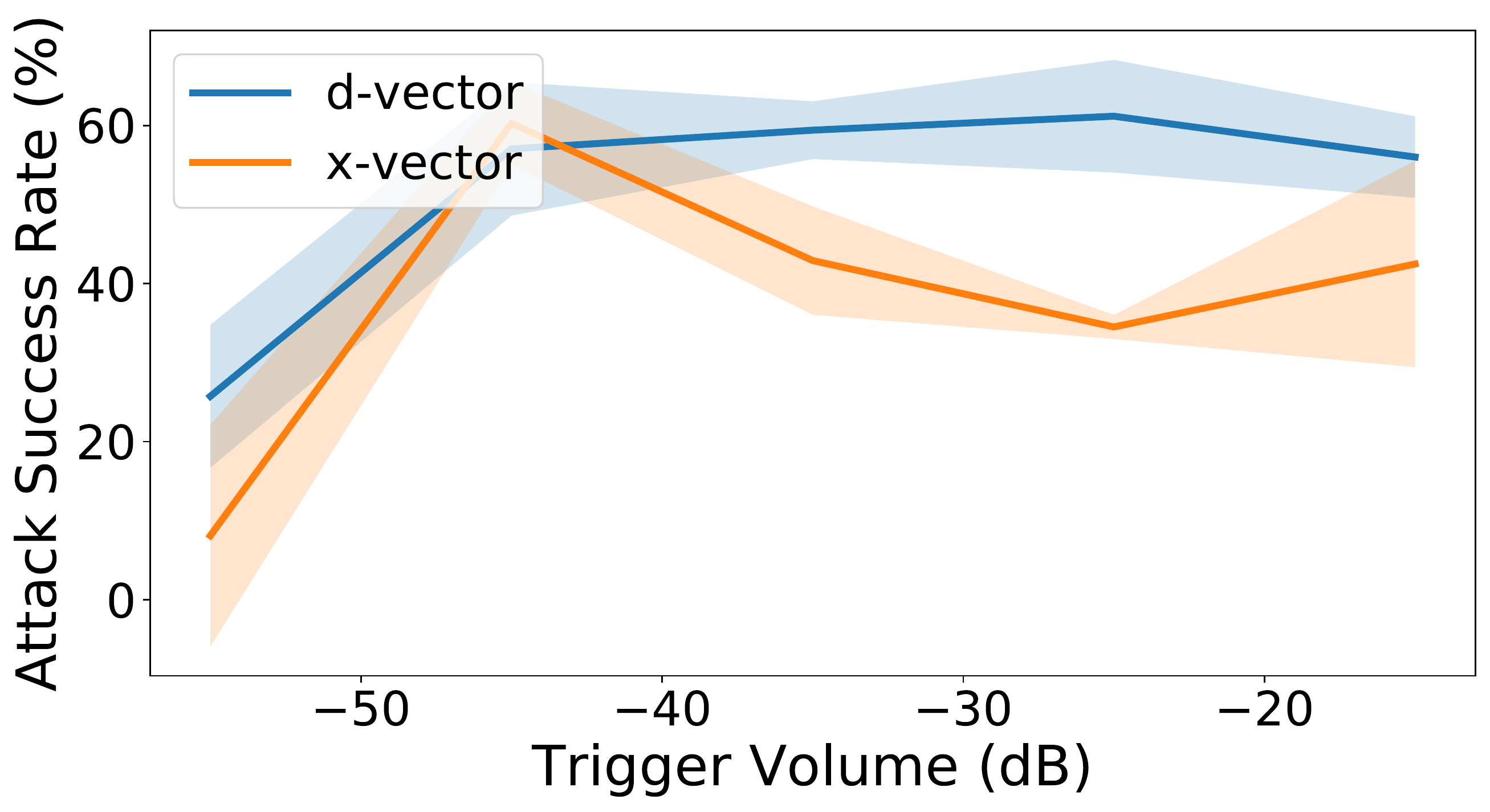}
}
\qquad\qquad
\subfigure{
\includegraphics[width=0.36\textwidth]{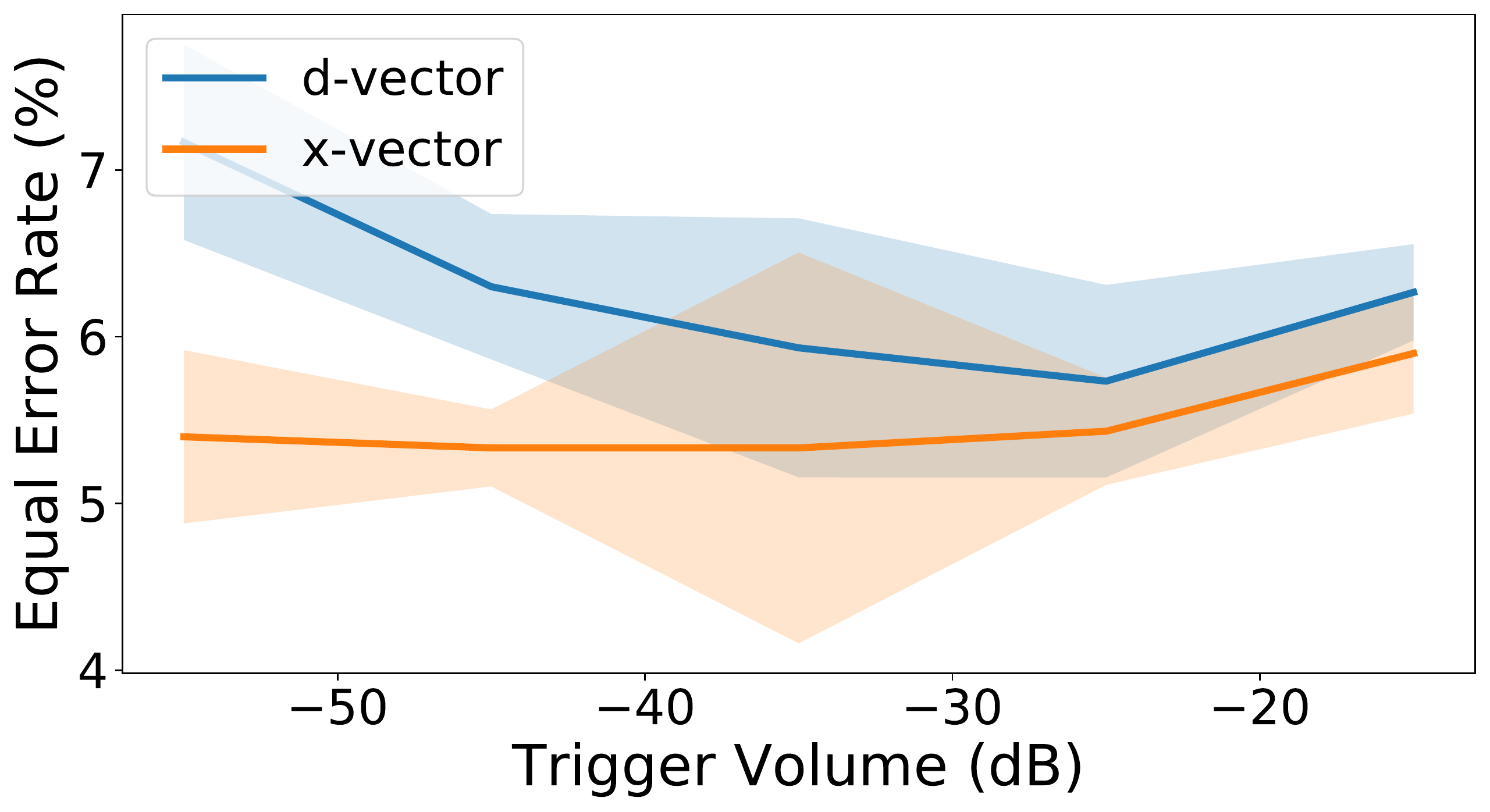}
}
\subfigure{
\includegraphics[width=0.36\textwidth]{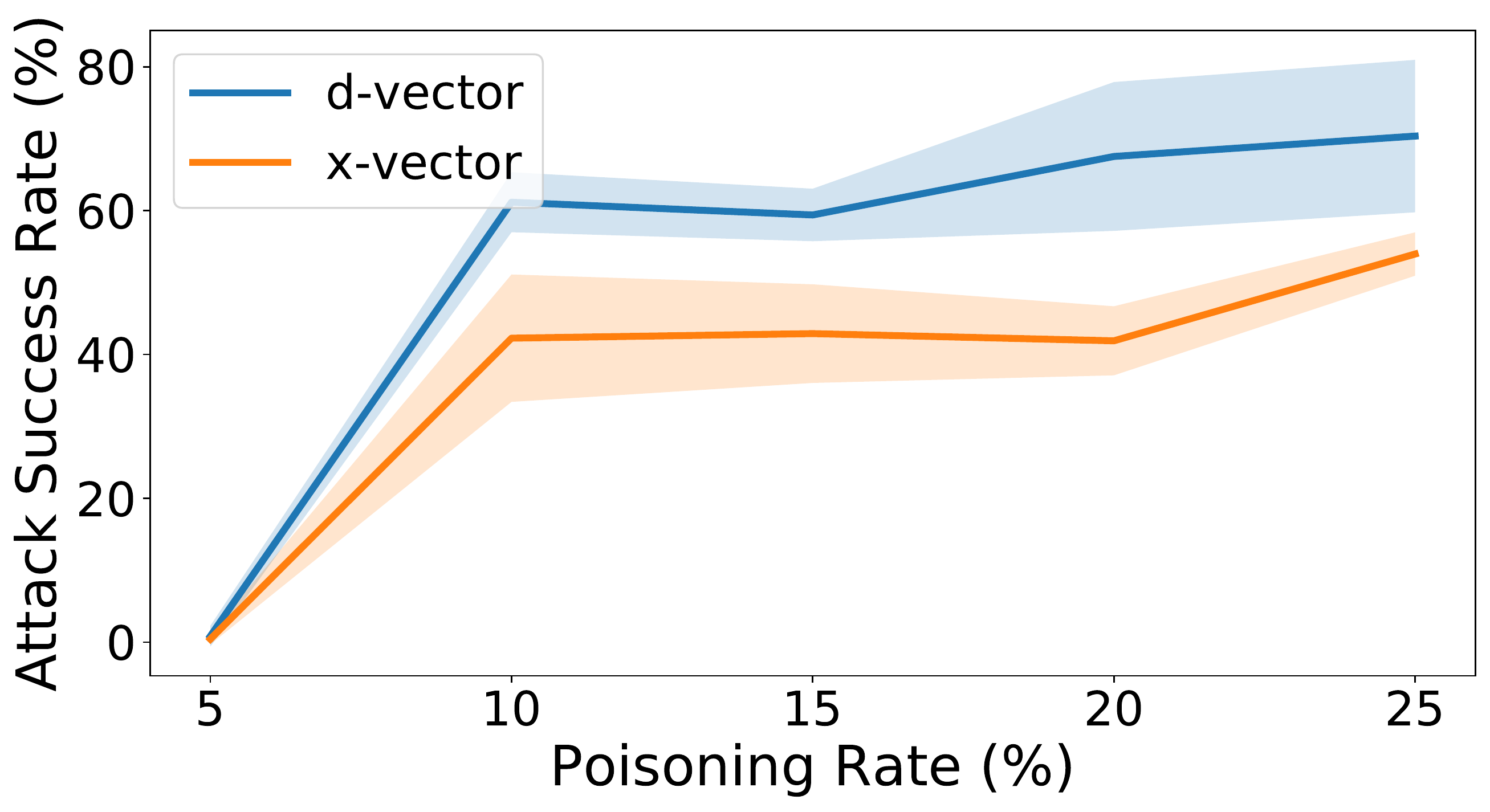}
}
\qquad\qquad
\subfigure{
\includegraphics[width=0.36\textwidth]{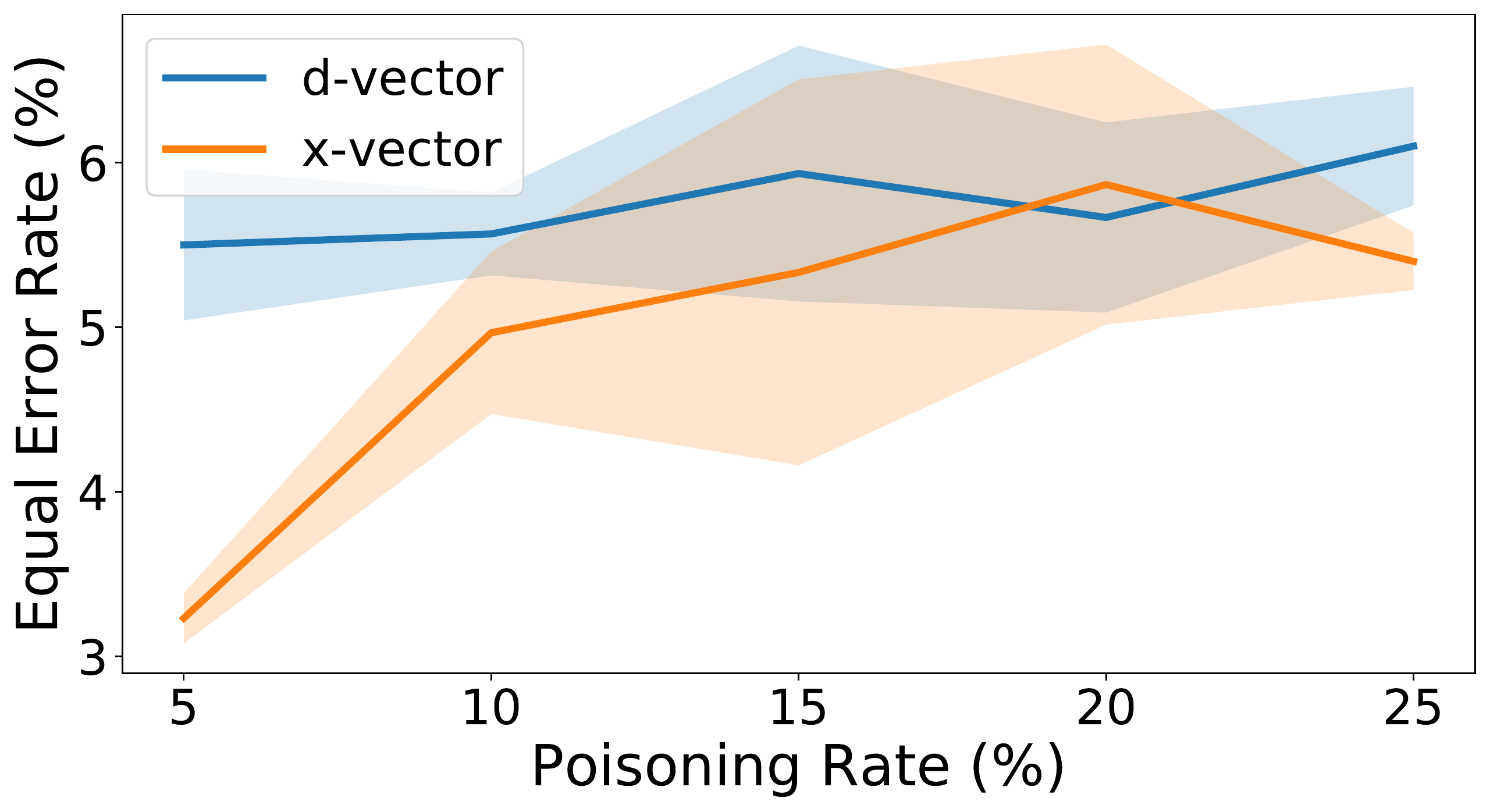}
}
\vspace{-0.4em}
\caption{The ASR (\%) and EER (\%) $w.r.t.$ different hyper-parameters on the TIMIT dataset. The background color indicates the standard deviation over all experiments.}
\label{Fig_abla}
\vspace{-1.2em}
\end{figure*}

\begin{table}[ht]
\centering
\caption{The EER (\%) and ASR (\%) of attacks on the TIMIT and VoxCeleb dataset. `EER' and `ASR' indicate the equal error rate and the attack success rate, respectively. The boldface indicates results with the best attack performance.}
\begin{tabular}{c|c|cc|cc}
\hline
\multicolumn{2}{c|}{Dataset $\rightarrow$}         & \multicolumn{2}{c|}{TIMIT} & \multicolumn{2}{c}{VoxCeleb} \\ \hline
Model $\downarrow$                     & Attack $\downarrow$   & EER         & ASR          & EER        & ASR        \\ \hline
\multirow{3}{*}{d-vector} & Benign & 4.3         & 2.5          & 12.0       & 4.0       \\ 
                          & BadNets  & 7.7         & 0.0          & 21.1      & \textbf{99.5}       \\
                          & ours     & 5.3         & \textbf{63.5}          & 13.0       & 52.0       \\ \hline
\multirow{3}{*}{x-vector} & Benign & 3.1         & 0.0          & 11.3       & 0.8        \\
                          & BadNets  & 4.0         & 0.0          & 19.7       & 2.5        \\
                          & ours     & 4.0         & \textbf{47.3}         & 15.7       & \textbf{45.0}       \\ \hline
\end{tabular}
\label{tab:main}
\vspace{-0.6em}
\end{table}

Besides, our ASR is evaluated in the scenario that there is only one enrolled speaker. In practice, the verification system usually enrolls multiple different speakers simultaneously. In this case, the ASR of our method may be further improved. It will be discussed in our future work.

\subsection{Ablation Study}
In this section, we discuss the effect of hyper-parameters in our attack. Each experiment is repeated three times to reduce the effect of randomness. Except for the studied parameter, other settings are the same as those used in Section \ref{sec: main}.

\vspace{0.3em}
\noindent \textbf{Effects of the Number of Clusters. } As shown in the first row of Figure \ref{Fig_abla}, the ASR increases with the increase of the number of clusters in general. An interesting phenomenon is that the EER does not significantly change $w.r.t.$ the number of clusters. It indicates that attackers can obtain a better attack performance by increasing the number of clusters.

\vspace{0.3em}
\noindent \textbf{Effects of the Trigger Volume. } Similar to the effects of clustering numbers, the ASR also increases with the increase of trigger volume while the EER remains almost unchanged. Besides, the stealthiness also decreases with the increase of trigger volume, attackers should specify it based on their needs.

\vspace{0.3em}
\noindent \textbf{Effects of the Poisoning Rate. } As shown in the third row of Figure \ref{Fig_abla}, both ASR and EER are directly related to the poisoning rate. Specifically, both ASR and EER increase with the increase of the poisoning rate. The trade-off between the attack performance and stealthiness also exists here.

\section{Conclusion}

In this paper, we explored how to conduct the backdoor attack against speaker verification methods. Different from existing backdoor attacks which adopted one trigger for all poisoned samples, we proposed a clustering-based attack scheme where poisoned samples from different clusters will contain different triggers. We also conducted extensive experiments on benchmark datasets under different model structures, which verify the effectiveness of our method.

\newpage

\bibliographystyle{IEEEbib}
\bibliography{ref}

\end{document}